\documentstyle[prl,aps,multicol,amssymb]{revtex}
\begin{document}

\title{Raman Solitons and Raman spikes}
\author{ J. Leon and A.V. Mikhailov$^+$ \\
{\em Physique Math\'ematique et Th\'eorique, CNRS - UM2,
34095 MONTPELLIER (France)\\
${ }^+$ Landau Institute for Theoretical Physics, MOSCOW (Russia) and\\
Dept. Applied Math., Leeds University, LS2 9JT, LEEDS (U.K.)}}
\maketitle


\begin{abstract}
Stimulated Raman scattering of a laser pump pulse seeded by a Stokes pulse
generically leaves a two-level medium initially at rest
in an excited state constituted of {\em static solitons} and radiation. 
The soliton birth manifests as sudden very large variations of the phase
of the output pump pulse.
This is proved by building the IST solution of SRS on the semi-line, 
which shows moreover that initial Stokes phase flips induce 
Raman spikes in the pump output also for short pulse experiments.
\end{abstract}

\begin{multicols}{2}

\paragraph*{Introduction.}

Stimulated Raman scattering (SRS) of high energy Laser pulse in two-level
medium has been intensively studied these last years
(see e.g. the review \cite{raywal}), more
especially after the experiments by Dr\"uhl, Wenzel and Carlsten
\cite{druwen} showing that spikes of pump radiation (Raman spikes)
occur spontaneously in the pump depletion zone, see also \cite{gakhov}.
Consequently nonlinear Raman amplification was used as a means to observe
at the macroscopic level the fluctuation of the phase of the initial
Stokes vacuum \cite{stokes}.

The SRS equations possess a Lax pair \cite{chusco} and, as a consequence, 
theoretical and experimental works were partly devoted to the search of
the {\em Raman soliton} predicted in \cite{chusco}. In particular
it has been believed \cite{soliton} that the {\em spike of pump radiation}
observed in  \cite{druwen} is a soliton which, in the 
inverse spectral transform scheme (IST), 
would be related to the discrete part of the spectrum.
However it has been proved that the observed Raman spike is not a soliton
but merely a manifestation of the continuous spectrum \cite{srs-leon}.

Then the  question of the creation a genuine Raman soliton is still an
open  problem (for 20 years now). We show here that the SRS
equations, solved as a boundary value problem on the semi-line, do 
induce the generation of solitons by pair, and that, after the passage of
the  pulses, the solitons are static in the medium.  Although the solitons
are not seen in the output pump {\em intensity} (where only Raman spikes
are seen), the signature of the soliton birth consists of very large
fluctuations of the {\em phase} of the pump output, which remains to be
experimentaly  observed.

Last we prove that the Raman spikes, observed in long pulse experiments 
(pulses with fwhm of 75 ns in \cite{druwen} and 15 ns in \cite{gakhov}),
where {\em damping} is essential in the model, must occur
also for short pulses (no damping), and give the condition
for their creation.

The efficiency of the method of solution serves not only as a basis for
seeking the Raman soliton, but also to evaluate {\em exactly} the result 
of any Raman amplification. Indeed the SRS equations are solved for {\em
arbitrary} values of the input ($x=0$) pump and Stokes pulses on a medium
with any initial ($t=0$) state. The original problem is reduced to a
Riccati equation (\ref{rho-evol}) and linear integral equations
(\ref{cauchy}). In many interesting cases the Riccati equation
(\ref{rho-evol}) can be explicitly solved (see (\ref{rho-sol})) and
important physical information can be obtained without solving equations
(\ref{cauchy}), such as the output (\ref{pump-out-infty}).

\paragraph*{The SRS equations.}

For a medium of length $L$, a detuning  $\Delta\omega$  and a 
phase missmatch $\Delta k$, in the dimensionless variables
$x=Z/L,\quad t= (cT-\eta Z)/L$, where $Z$ and $T$ are the physical
variables, the SRS equations for short pulses (infinite dephasing time) 
in the {\em slowly varying envelope approximation} read \cite{yariv}
\begin{eqnarray}\label{maxsimp}
\partial_x a_L&&=q  a_S\ e^{-i\Delta\omega t}\ e^{2ikx},\nonumber\\
\partial_x a_S&&=-\bar q a_L\ e^{i\Delta\omega t}\ e^{-2ikx},\\
\partial_tq &&= - g \int dk  a_L a_S^*\ e^{i\Delta\omega t}\
e^{-2ikx}.\nonumber
\end{eqnarray}
The phase missmatch is represented there by the dimensionless parameter
$k=\frac 12 L(\Delta k-\frac\eta c\Delta\omega)$. The envelopes $a_L$ 
(frequency $\omega_L$) and $a_S$ (frequency $\omega_S$) 
are scaled to the peak intensity $I_m$ of the input laser field
and the dimensionless coupling constant is given by
\begin{equation}
g =\frac{L^2}{16}{N\alpha'_0\epsilon_0\over\eta mc^2}
{\omega_S\over\omega_V}\ I_m^2,
\label{qdyn}\end{equation}
where $\alpha'_0$ is the differential
polarizability at equilibrium, $c/\eta$ is the light velocity in the
medium of transition frequency $\omega_V$, and $N$ is the density 
of oscillators of mass $m$.
We have considered in (\ref{maxsimp}) the cooperative interaction of all
$k$-components with the medium and then the input ($x=0$) values of the
light waves are also function of $k$ sharply distributed around $k=0$ 
\begin{equation}
a_L(k,0,t)=J_L(k,t),\quad a_S(k,0,t)=J_S(k,t).
\label{in-A}\end{equation}
Finally the electromagnetic field is obtained as
$$E=\frac{I_m}2
a_L e^{i(k_LZ-\omega_LT)}+
\frac{I_m}2\sqrt{\frac{\omega_S}{\omega_L}}
a_Se^{i(k_SZ-\omega_ST)}+c.c.$$

The system (\ref{maxsimp}) can be solved on the infinte line for 
arbitrary asymptotic boundary values \cite{gabzak} and it
has been proved in \cite{srs-leon} that this allows the
interpretation of the experiments of \cite{druwen}. We will demonstrate 
that it can be solved completely on the semi-line $x\in[0,\infty)$, which
actually furnishes the solution at any point $x=L$ and hence solves the
finite  interval case as a {\em free end problem}.

\paragraph*{Lax pair.}

The time evolution in (\ref{maxsimp}) results as the compatibility 
condition $U_t-V_x+[U,V]=0$ for the Lax pair ($x\ge0$, $t\ge0$)
\begin{eqnarray}
\varphi_x &=& U \varphi+ik\varphi \sigma_3,
\label{lax-pair-x}\\
\varphi_t &=& V \varphi+
\varphi e^{-ik\sigma_3x}\Omega e^{ik\sigma_3x},
\label{lax-pair-t}\end{eqnarray}
where the {\em dispersion relation} $\Omega$ is $x$-independent and
\begin{eqnarray}
U&=&\left(\matrix{-ik&q \cr -\bar q &ik}\right),
\nonumber\\
V&=&\frac{g }{4i}\int\frac{d\lambda}{\lambda-k}
\left(\matrix{|a_L|^2-|a_S|^2 & 2a_L\bar a_Se^{-2i\lambda x}\cr
2\bar a_La_Se^{2i\lambda x}   &|a_S|^2-|a_L|^2}\right).
\end{eqnarray}
$V(k)$ is discontinuous, therefore only the limits
$V^\pm=V(k\pm i0)$ are defined on the real $k$-axis. 
Such is also the case for $\Omega$, and $\Omega^\pm$
will have to be determined from the boundary conditions on the solution
$\varphi$.

\paragraph*{Scattering problem.}

Following standard methods in spectral theory,
see e.g. \cite{levitan}, we define the solutions $\varphi^+$ and 
$\varphi^-$ according to
\begin{eqnarray}\label{voltera}
\left(\matrix{\varphi_{11}^+ \cr\varphi_{21}^+}\right)&=&
\left(\matrix{1\cr0}\right)-
\left(\matrix{-\int_{0}^{x}d\xi  q \varphi_{21}^+ \cr
      \int_{0}^{x}d\xi  \bar q \varphi_{11}^+ e^{2ik(x-\xi )}}
       \right)\nonumber\\                          
\left(\matrix{\varphi_{12}^+ \cr\varphi_{22}^+}\right)&=&
\left(\matrix{0\cr1}\right)-
\left(\matrix{\int_{x}^{\infty}d\xi  q \varphi_{22}^+ 
                              e^{-2ik(x-\xi )}\cr
     \int_{0}^{x}d\xi \bar q \varphi_{12}^+}
       \right)\nonumber\\  
\left(\matrix{\varphi_{11}^- \cr\varphi_{21}^-}\right)&=&
\left(\matrix{1\cr0}\right)+
\left(\matrix{\int_{0}^{x}d\xi  q \varphi_{21}^- \cr
     \int_{x}^{\infty}d\xi\bar q \varphi_{11}^- e^{2ik(x-\xi )}}
       \right)\nonumber\\ 
\left(\matrix{\varphi_{12}^- \cr\varphi_{22}^- \cr}\right)&=&
\left(\matrix{0\cr1}\right)+
\left(\matrix{\int_{0}^{x}d\xi  q \varphi_{22}^- 
                              e^{-2ik(x-\xi )}\cr
     -\int_{0}^{x}d\xi \bar q \varphi_{12}^- } \right).
\end{eqnarray}
These solutions obey the {\em reduction} $\bar\varphi_1^+(\bar k)
=i\sigma_2\varphi_2^-(k)$, the Riemann-Hilbert relations
\begin{eqnarray}\label{R-H}
\varphi_1^- &=&\varphi_1^+ -e^{2ikx}\rho^*\varphi_2^-,\nonumber\\
\varphi_2^+ &=& \varphi_2^- + e^{-2ikx}\rho\varphi_1^+, 
\end{eqnarray}
and the bounds
\begin{eqnarray}\label{bound-x}
x=0\ : && \varphi^+=\left(\matrix{1 & \rho \cr 0 & 1}\right),\quad
\varphi^-=\left(\matrix{1 & 0 \cr -\rho^*  & 1}\right),
\nonumber\\
x\to\infty\ : &&\varphi^+\to
\left(\matrix{1/\tau & 0\cr e^{2ikx}\rho^*/\tau^* &\tau}\right),
\nonumber\\
&&\varphi^-\to\left(\matrix{\tau^*&-e^{-2ikx}\rho/\tau\cr 
0&1/\tau^*}\right).
\end{eqnarray}
which define the reflection coefficient $\rho$ and
the transmission coefficient $\tau$ ($\rho^*(k)$ stands for
$\bar\rho(\bar k)$). Note that $\det\{\varphi(x)\}=
\det\{\varphi(\infty)\}=1$, hence for real $k$  $|\tau|^2=1+|\rho|^2$.

The  column vectors $\varphi_1^+$ and $\varphi_2^-$  are entire functions
in the $k$-plane, with good behaviors as $k\to\infty$ in the upper
half-plane for $\varphi_1^+$ (and the lower one for $\varphi_2^-$).
The vector $\varphi_2^+$ is meromorphic in the upper half-plane
with a finite number $N$ of simple poles $k_n$ (and $\varphi_1^-$ in
the lower one with poles $\bar k_n$). Consequently $\tau$ and $\rho$
have meromorphic extensions in the upper half-plane where they possess the
$N$ simple poles $k_n$ (the bound states locations).

\paragraph*{Solution of SRS.}

 From the boundary data (\ref{in-A}) we have readily
(set $\Delta \omega =0$ for simplicity because
it can be scaled off in $a_S$)
\begin{equation}\label{exp-vect}
\left(\matrix{a_L\cr a_Se^{2ikx}}\right)=
J_L\ \varphi_1^+ +J_S\ \varphi_2^- e^{2ikx}.
\end{equation}
After careful computations of the boundary values of (\ref{lax-pair-t}), 
using separately $\varphi^+$ and  $\varphi^-$,
we obtain both limits $\Omega^+$ and $\Omega^-$ of $\Omega$ 
and the following time evolution of the scattering
coefficient $\rho(k,t)$:
\begin{eqnarray}\label{rho-evol}
&&\rho_t=-\rho^2\ {\cal C}^+[m^*]-2\rho\ {\cal C}^+[\phi]-{\cal C}^+[m].
\\
&&{\cal C}^+[f]=\frac1\pi\int\frac{d\lambda}{\lambda-(k+i0)}\ f(\lambda),
\nonumber\\
&&m=\frac{i\pi}{2}g J_LJ_S^*,\quad \phi=\frac{i\pi}{4}g (|J_L|^2-|J_S|^2).
\nonumber\end{eqnarray}

The reconstruction proceeds through the solution of the Riemann-Hilbert
problem (\ref{R-H}) which reads 
\begin{eqnarray}\label{cauchy}
\varphi_1^+(k)=\left(\matrix{1\cr0}\right)+&&{1\over2i\pi}
\int_{C_-}{d\lambda\over\lambda-k}
\rho^*(\lambda)\varphi_2^-(\lambda) e^{2i\lambda x},\nonumber\\
\varphi_2^-(k)=\left(\matrix{0\cr1}\right)+&&{1\over2i\pi}
\int_{C_+}{d\lambda\over\lambda-k}
\rho(\lambda)\varphi_1^+(\lambda) e^{-2i\lambda x},
\end{eqnarray}
where $C_+$ (resp. $C_-$) is a contour from $-\infty+i0$ to $+\infty+i0$
(resp. from $-\infty-i0$ to $+\infty-i0$) passing over 
all poles $k_n$ of $\rho(k)$ (resp. under $\bar k_n$). 
Remember the notation $\rho^*(\lambda)=\bar\rho(\bar\lambda)$.

Since $m$ and $\phi$ are continuous bounded functions of $k$, the
coefficients of the Riccati equation (\ref{rho-evol}) are analytic
functions in the upper half $k$-plane, where precisely $\rho$ is
meromorphic at time $t=0$. Consequently $\rho$ remains meromorphic at
later times, which allows to prove that the
reconstructed fields do obey the system of equations  (\ref{maxsimp}),
or in short that the reconstructed $\varphi$ does obey (\ref{bound-x}).
Indeed from (\ref{cauchy}), the Cauchy theorem leads to
\begin{equation}
\forall x<0 \quad : \quad
\varphi_1^+=\left(\matrix{1\cr0}\right),\quad
\varphi_2^-=\left(\matrix{0\cr1}\right).
\label{val-0-1}\end{equation}

The formulae (\ref{rho-evol}) and (\ref{cauchy}) with (\ref{exp-vect})
furnish the solution of (\ref{maxsimp}), in particular the output field
values at $x=L$ (whatever be $q$ for $x>L$). In the limit $L\to\infty$,
and  by (\ref{bound-x}), the solution becomes {\em explicit}, and for
instance the output pump reads
\begin{equation}\label{pump-out-infty}
x\to\infty \ :\ a_L\to\frac1\tau J_L-\frac\rho\tau J_S.\end{equation}

\paragraph*{Evolution of laser pulses.}

We are interested here in the physically relevant case when the medium is
initially in the groud state, that is 
$\rho(k,0)=0$.
Next we choose a Stokes wave as a portion of the pump wave, both with
Loretzian lineshape, namely, for $\kappa>0$,
$$|J_L|^2=|A(t)|^2 \ \frac{\kappa}{\pi (k^2+\kappa ^2)},\quad
J_{S}=e^{-\gamma-i\theta(t)} J_{L}$$
where $A(t)$ is the pulse shape of duration $t_m$
($t>t_m\Rightarrow A(t)=0$). The parameter $\gamma$ (real 
positive constant) measures the ratio pump/Stokes inputs. Considering the 
case when the input Stokes wave experiences one phase flip at some time
$t_0$, we choose the phase $\theta(t)=0$ for $t<t_0$ and $\theta(t)=\pi$ 
for $t>t_0$.

It follows from contour integration
that the Riccati equation (\ref{rho-evol}) can be rewritten as
$$\rho_t=-\frac i2 \frac{ge^{-\gamma}}{k+i\kappa}|A|^2
\left[\rho^2e^{-i\theta}-2\rho\sinh\gamma-e^{i\theta}\right].$$
Taking into account the phase flip,
the general solution with zero initial datum reads
\begin{eqnarray}\label{rho-sol}
t\le t_0 &:\ & \rho=\frac{\sinh\delta}{\cosh(\delta-\gamma)},\\
t\ge t_0 &:\ &
\rho=\frac{\rho_0\cosh(\delta-\delta_0+\gamma)-\sinh(\delta-\delta_0)}
{\cosh(\delta-\delta_0-\gamma)-\rho_0\sinh(\delta-\delta_0)},
\nonumber\end{eqnarray}
with $\rho_0=\rho(k,t_0),\ \delta_0=\delta(k,t_0)$ and the definitions
\begin{equation}\label{delta}
\delta(k,t)=\frac{i T(t)}{k+i\kappa},\quad
 T(t)=\frac14g(1+e^{-2\gamma})\int_0^tdt'|A|^2.
\end{equation}

\paragraph*{Soliton generation and signature.}

Let us consider first the case with no phase flip, that is $t_0>t_m$.
The spectral transform $\rho$, first function in (\ref{rho-sol}),
has an infinite set of moving
poles $k_n(t),\ n\in\mathbb Z$, given by
\begin{equation}\label{pole}
k_n=-i\kappa+\frac{T(t)}{(n+\frac12)\pi-i\gamma},
\end{equation}
which are associated with solitons as soon as they lie in the upper half
plane (in the lower half plane they are the resonances). 

As t evolves, and for a given linewidth $\kappa$, these poles move
from the point $-i\kappa$ and they may cross the real axis (and generate
solitons) if $T(t)$ is large enough, which means enough energy in the
input pulses. Moreover, since $k_n=-\bar k_{-n-1}$, solitons are created
by pair. After the passage of the pulse,
($t>t_m$), $m$ and $\phi$ vanish in (\ref{rho-evol}), $T(t)$ is constant,
and hence the whole solution becomes $t$-independent.
Consequently the laser pulses leave in the medium a {\em finite number
of static bi-solitons}. 

The question of their observation is not straightforward. Indeed,
taking for simplicity the case of an infinite medium, the pump
output is given in (\ref{pump-out-infty})
where the coefficients $\rho$ and $\tau$ are understood for real values of
$k$ (the essential missmatch). The main point here is that both 
$1/\tau$ and $\rho/\tau$ are holomorphic functions
ion the upper half-plane of $k$, and hence an 
integrated intensity like $\int dk|a_L|^2$ will not show the presence of
the poles $k_n$. Moreover, if at $t=t_s$ a resonance crosses the real axis 
at $k_s$ to become a bound state (soliton birth), the output 
{\em intensity} at that particular value $k_s$ will simply show a full
depletion ($1/\tau(k_s)=0$ and $|\rho(k_s)/\tau(k_s)|=1$).

However, the {\em phase} of the transmission coefficient $\tau(k)$
diverges  for all real $k$ precisely at $t=t_s$. Indeed the {\em trace
formula} (with no pole on the real axis i.e. just before pole crossing) 
reads for $k\in \mathbb R$ \cite{book}
$$\frac{\tau(k)}{|\tau(k)|}=\prod{k-\bar k_n\over k-k_n}
\exp[-{i\over2\pi}\int{\ln(1+|\rho(\lambda)|^2)\over\lambda-k}],$$
where the integral over $\lambda$ is understood as the Cauchy principal
value, and where the product concerns only the bound states if any. 
It is then clear that,
for every $k\ne k_s$, the phase of $\tau(k)$ diverges as $t\to t_s$
because $\rho(\lambda)$ has a pole in $\lambda=k_s$. The observation of
true Raman solitons should go then with a measure of the phase of the pump 
output.

We remark here that, in the {\em sharp line limit} $\kappa\to 0$ and with
absense of a flip in the boundary condition, the infinite set of poles
$k_n$ lie in the upper half-plane and $|k_{n}|$ is proportional to $T(t)$
(\ref{pole}), which implies that the solution becomes
self-similar (the self-similar solution of SRS is
analized in \cite{winter}).

\paragraph*{Stokes phase flip and Raman spike.}

As shown in \cite{srs-leon}, the Raman spike observed in \cite{druwen} and
largely studied later, occurs at time $t_r$ for which $\rho(k,t_r)=0$.
Indeed, then $|\tau(k,t_r)|=1$ and from (\ref{pump-out-infty}) the pump 
(intensity) is fully repleted.

The second expression in (\ref{rho-sol}) gives that $\rho$
vanishes at $t_r$ for
\begin{equation}\label{spike-semi}
\coth(\delta_0)-\coth(\delta_r-\delta_0)=2\tanh\gamma,
\end{equation}
which relates $t_0$ (instant of phase flip) and $t_r$ (instant of Raman
spike) to the pulse area through (\ref{delta}). 

To get a simple insight in that formula, we
may consider the dominant term $k=0$ (resonance) for which $\delta_0$ 
and $\delta_r-\delta_0$ are real real positive. Then a Raman spike will
be allowed if (at $k=0$)
$$\coth(\delta_m-\delta_0)>\coth(\delta_0)-2\tanh\gamma>0,$$
where $\delta_m=\delta(k,t_m)$, and $t_m$ is the pulse duration.
This is obtained without damping, which suggests
the search of Raman spikes in short pulse experiments
(short with respect to the medium dephasing time).

We note finally that the occurence of a phase flip alters the
denominator of $\rho$, which can be used to change the discrete spectrum
and hence to modify the soliton part of the solution.

\paragraph*{Conclusion.}

The following results have been obtained:\\
1 - The proof that genuine Raman solitons do occur in SRS experiments and
cannot be detected in the energy of the output but rather in its phase
divergence.\\
2 - The {\em spontaneous} generation of solitons out of vacuum
in  an integrable system.\\
3 - The complete solution by IST of the boundary value problem for SRS on
the semi-line. This solution relies on the analytical properties of the
nonlinear Fourier transform $\rho(k,t)$ which are {\em conserved} in the
time evolution.\\
4 - The proof that Raman spikes are {\em generic} as they occur as well for
short duration pulses, as suspected in the experimental work 
\cite{gakhov}.\\
5 - The proof  that in the sharp line limit and with the proportional
pump/Stokes input on a medium initially at rest, the solution is 
self-similar.

\paragraph*{Comments.}

1 - The IST solution of SRS  on the semi-line has been considered in
\cite{kaup}  for the sharp line case. Following the same procedure 
in our {\em inhomogeneously broadened} case, and
in particular neglecting some terms
in the spectral  data evolution, would lead to an evolution for $\rho$
with non-analytic coefficients, in contradiction with the necessary
meromorphy  of $\rho$.\\
2 - IST on the semi-line for the nonlinear Schr\"odinger equation
has been considered in \cite{fokas}, and the problem is considerably more
difficult mainly because of the necessity of the knowledge of additional
constraints in $x=0$ (e.g. the time derivative of the field).\\
3 - Boundary conditions for nonlinear evolution equations are known to be
a source for soliton generation \cite{chouchu} \cite{fokas}, but
this is the first instance of an integrable system which 
shows {\em explicitely} the spontaneous generation of solitons,
as motion of poles from the lower to the upper half-plane, from 
resonances to  bound states.\\
4 - At the moment we are not able to extract, from the explicit form of
the spectral data, the explicit form of the {\em potential} (the medium
excitation field) because on the semi-line no pure soliton exist
(would $\rho$ vanish on the real axis that it would vanish everywhere
and no bound states would be allowed).  
However the method provides the explicit form of the physically
relevant quantities which are the light pulses output values.\\
5 - The manifestation of the soliton birth as sudden large variation
of the phase of the output was quite unsuspected
and should lead to reconsider the measures of the output in SRS
experiments.

\paragraph*{\bf Aknowledgements.} 

This work has been done as part  of the programm
{\em Dynamique nonlin\'eaire et chaines mol\'eculaires - UM2}. 
A.V.M. is grateful to the {\em Laboratoire de Physique Math\'ematique et
Th\'eorique} for hospitality and to the CNRS for support.

\end{multicols}

\end{document}